\documentclass[12pt]{article}
\begin{document}
\begin{titlepage}

\author { VICTOR NOVOZHILOV \footnote{vnovo@vn9711.spb.edu} and
  YURI NOVOZHILOV \footnote{yunovo@pobox.spbu.ru} \\
 V.Fock Institute of Physics, St.Petersburg State University,\\
  Ulyanovskaya str. 1, 198504, St.Petersburg, Russia}
\title{\bf  COLOR BOSONIZATION, CHIRAL PARAMETRIZATION OF GLUONIC FIELD and QCD
EFFECTIVE ACTION}
\date{}
\maketitle

\abstract {We develop a color bosonization approach to treatment
of QCD gauge field (''gluons'') at low energies in order to derive
an effective color action of QCD taking into account the quark
chiral anomaly in the case of SU(2) color.. We have found that
there exists such a region in the chiral sector of color space,
where a gauge field coincides with of chirally rotated vector
field, while an induced axial vector field disappears. In this
region, the unit color vector of chiral field plays a defining
role, and a gauge field is parametrized in terms of chiral
parameters, so that no additional degrees of freedom are
introduced by the chiral field. A QCD gauge field decomposition in
color bosonization is a sum of a chirally rotated gauge field and
an induced axial-vector field expressed in terms of gluonic
variables. An induced axial-vector field defines the chiral color
anomaly and an effective color action of QCD. This action admits
existence of a gauge invariant d=2 condensate of induced
axial-vector field and mass.}

\vskip 6pt

Keywords: [QCD effective action, color chiral model]

\vskip 6pt

PACS Nos.: 12.38.Aw,12.39.Fe

\end{titlepage}

  {\large \bf 1. Introduction}

\vskip 9pt

It is widely believed that the key point to the infrared QCD and
confinement is to find important field configurations in the color
space. To this end, convenient parametrization of the gauge field
is considered as the first task. This task contains several subtle
points, because, due to the phenomenon of chiral anomaly
\cite{ABJ}, the total color space unifies gauge and chiral
(anomalous) sectors in the unique sector with topological
properties. The gauge sector was extensively studied in pursuite
of the monopole condensation scenario for confinement
\cite{nambu}. Recently, color Hopf-type solitons were reported by
Faddeev and Niemi \cite{Faddeev+Niemi} on the basis of n-field
model of infrared QCD by Faddeev \cite{Faddeev75}. This result
generated a series of papers explaining the place of the n-model
in infrared QCD \cite{R.Battye and P.Sutcliffe},\cite{S.Shabanov},
its relation to the Cho decomposition of the QCD gauge field
\cite{Cho, Cho+Kimm}, as well as transformation properties
\cite{kubo}. In quest for infrared variables in QCD two versions
of parametrization of the gauge field in terms of dual variables
were proposed by Faddeev and Niemi \cite{Faddeev-02}. The chiral
color sector at low energies was studied less. The color chiral
bosonization \cite{A+N}describes this sector in terms of an
effective action, and it was found that such an action with a
chromomagnetic vacuum background field admits stable soliton -
skyrmion solutions \cite{N+N}.

In the present paper we develop chiral color bosonization approach
to low energy QCD in combined gauge and chiral sectors, when quark
color chiral degrees of freedom are explicitly taken into account.
The aim of the paper is to find the Lagrangian describing the
quark color chiral contribution to QCD dynamics.

 We assume that bosonization is possible in the color case
and introduce local chiral color transformation of quarks, thereby
extending quark color group to the Left-Right group and including
the chiral color anomaly into consideration. This approach has an
advantage of well defined initial point, i.e. two separate color
sectors - gluons and chirally rotated quarks - with definite
transformation properties (Left-Right color group). We have an
initial gauge field $V_\mu $ and two fields arising in its chiral
transformation $U$, namely, a gauge field $V_\mu ^U$ and an axial
vector field $A_\mu ^U$.However, introducing the quark color
chiral field $U$ we arrive at a system with too many degrees of
freedom. A consistent parametrization of gluons and the chiral
field within Left-Right color group becomes a special task. To
eliminate superfluous variables, one should consider a relation
between gluonic fields $V_\mu $ and chirally transformed field
$V_\mu ^U$ and find how different parts of these fields can be
made from the same material, so that chiral field variables are
either fixed, or fully incorporated into gauge field variables.
This is a key point in our color bosonization approach. When this
task is accomplished, one can integrate chiral color anomaly and
get an effective action, as a sum of bosonization action and the
Yang-Mills action. We hope that this effective action will be
helpful in low energy QCD.

We consider gluons and quarks at one-loop level in the background gauge.
Then gauge fields $V_\mu ,V_\mu ^U$ and an axial vector $A_\mu ^U$ in the
Dirac Lagrangian are background ones. One-loop quark chiral terms represent
bosonization corrections to the Yang-Mills action for the background gauge
fields. One-loop gluon ic action is of no interest for the present paper.

In order to find, how vectors $\left( V^U,A^U\right) $ are
expressed in terms of gluonic field $V_\mu $ variables, we look at
first for a region $\Omega $ in color space and such a gauge field
there $V_\mu \equiv V_\mu ^\Omega $, which is invariant under
chiral transformation: ($V_\mu ^\Omega )^U=V_\mu ^\Omega $. Such a
gauge field $V_\mu ^\Omega $ will contain $U$ -variables. With
$V_\mu =V_\mu ^\Omega $, an axial vector field should be absent
$A_\mu ^\Omega =0$ and quark color chiral anomaly disappear. This
case provides us with a partial parametrization of gluonic field
in terms of chiral parameters. In order to investigate this point
in the classical theory with background fields we need a relation
expressing gluonic field $ V_\mu $ in terms of chirally dependent
field $V_\mu ^U\left( x\right) $, or the determinant $\det \left(
\partial V/\partial V^U\right) $. In quantum theory we need to
know how changes the measure $d\mu _V\rightarrow d\mu _V^U$ of the
gluonic path integral because of replacement $V_\mu \left(
x\right) \rightarrow V_\mu ^U\left( x\right) $.

We develop the theory at example of the color SU(2) group. An
extension to the color SU(3) will use many points of the SU(2)
case \cite{victor}.

\vskip 9pt

 {\large \bf 2. Bosonization in the flavor case}

\vskip 9pt

In this section we review the bosonization approach to the
effective chiral action in the case of chiral flavor. We consider
massless fermions in external vector and axial vector fields
$V_\mu ,A_\mu $. The path integral of fermions $Z_\psi \left[
V,A\right] $ is a functional of $V,A$:

$$
Z_\psi \left[ V,A\right] =\int d\mu _\psi \exp i\int
dx\overline{\psi }\not D\left( V,A\right) \psi \eqno(1)
$$

where $\not D=i\gamma ^\mu \left( \partial _\mu +V_\mu +i\gamma
_5A_\mu \right) $ is the Dirac operator. The chiral transformation
of fermions is given by

$$
\psi _L^{\prime }=\xi _L\psi _L,\psi _R^{\prime }=\xi _R\psi
_R,\psi =\psi _L+\psi _R\eqno(2)
$$

where $\xi _L\left( x\right) $ and $\xi _R\left( x\right) $ are
local chiral phase factors of left and right quarks $\psi _L$ and
$\psi _R$, represented by unitary matrices in defining
representations of left $SU\left( N\right) _L $ and right
$SU\left( N\right) _R$ subgroups of the chiral group $G_{LR}=$
$SU\left( N\right) _L\times SU\left( N\right) _R$. For $\psi
_L=\frac 12\left( 1+\gamma _5\right) \psi ,\psi _R=\frac 12\left(
1-\gamma _5\right) \psi $, generators $t_{La}$ and $t_{Ra}$ of
left and right subgroups of $G_{LR}$ can be written as
$t_{La}=\frac 14\left( 1+\gamma _5\right) \tau _a,t_{Ra}=\frac
14\left( 1-\gamma _5\right) \tau _a,\left[ t_{La,}t_{Rb}\right]
=0$, where $\tau _a,a=1,2,3$ are the Pauli matrices. Then quark
left and right chiral phase factors $\xi _L,\xi _R$ arise from
application of operators $\hat \xi _L=\exp (-it_{La}\omega
_{La}),\hat \xi _R=\exp (-it_{Ra}\omega _{Ra})$ to left and right
quarks $\psi _L$ and $\psi _R$. Vector gauge transformations
$g\left( x\right) $ are associated with $ t_a=t_{La}+t_{Ra}=\tau
_a/2$ , i.e. $g\left( x\right) $ has properties of the product
$\hat \xi _L\left( x\right) \hat \xi _R\left( x\right) $ of
identical left and right rotations, $\omega _L=\omega _R$ $=\alpha
$. The generator of purely chiral transformations $g_5\left(
x\right) $ is $t_{5a}=\gamma _5\tau _a/2=t_{La}-t_{Ra}$; thus,
$g_5\left( x\right) $ has properties of $\hat \xi _L\left(
x\right) \hat \xi _R^{+}\left( x\right) $ for $\omega _L=\omega
_R=\Theta $. Infinitesimally, the Dirac operator is transformed
according to

$$
\delta \not D=[i\frac 12\alpha _a\tau _a,\not D]+\{i\frac 12\gamma
_5\Theta _a\tau _a,\not D\} \eqno(3)
$$

Commutation relations for $t_a,t_{5a}$ are given by

$$
\left[ t_a,t_b\right] =i\varepsilon _{abc}t_c,\left[
t_a,t_{5b}\right] =i\varepsilon _{abc}t_{5c},\left[
t_{5a},t_{5b}\right] =i\varepsilon _{abc}t_c \eqno(4)
$$

Instead of phases $\xi _L$ and $\xi _R$ one can work with the
chiral field $U=\xi _R^{+}\xi _L$, which describes rotation of
only left quark leaving right quark in peace $\psi _L\rightarrow
\psi _L^{\prime }=U\psi _{L,}\psi \rightarrow \psi _R^{\prime
}=\psi _R$. The same result can be obtained by the chiral
transformation $\psi _L\rightarrow \xi _L\psi _L,\psi
_R\rightarrow \xi _R\psi _R$, followed by a vector gauge
transformation with a gauge function $\xi _R^{+}$. The usual
chiral gauge choice is $\xi _R=\xi _L^{+}$, then the chiral field
is taken as squared left chiral phase: $U=\xi _L^2$.

The chiral tranformation of fermions in the Dirac action is equivalent to
the following change of the Dirac operator

$$ \bar \psi ^{\prime }\not D\left( V,A\right) \psi ^{\prime } =\bar
\psi D\left( V^U,A^U\right) \psi $$
$$ V^U=\frac 12[U_{}^{+}(\partial +V+A)U+(V-A)]  $$
$$ A^U=\frac 12[U^{+}(\partial +V+A)U-(V-A)] \eqno(5)$$

A transformed fermionic path integral $Z_\psi ^{\prime }$ because
of $\psi \rightarrow \psi ^{\prime }$ in the Dirac action is equal
to an original path integral as a functional of transformed fields
$Z_\psi ^{\prime }\left[ V,A\right] =Z_\psi \left[ V^{\prime
},A^{\prime }\right] $. $Z_\psi $ is invariant under vector gauge
transformations of fermions, but undergoes changes under chiral
tranformations, because of non-invariance of the fermionic measure
$d\bar \psi d\psi $ \cite{fujikawa}: chiral transformations are
anomalous. The chiral anomaly ${\cal A}$ is defined by an
infinitesimal change of $\ln Z_\psi $ due to an infinitesimal
chiral transformation $\delta g_5=i\theta _a\tau _a\equiv \Theta
$.

We put $g_5(s)=\exp \gamma _5\Theta s$ and write the anomaly
${\cal A}\left( x,\Theta \right) $ at a chiral angle $\Theta $

$$
{\cal A}\left( x,\Theta \right) =\frac {1}{i}
 \frac {\delta \ln Z_\psi \left(\exp \Theta s \right)}{\delta s}_{s=1}
  \eqno(6)
$$

The usual way \cite{AANN} to calculate effective chiral action
$W_{eff}$ is to find the anomaly and integrate it over $s$ up to
$g_5=\exp \gamma _5 \Theta $.

$$
W_{eff}^{}=-\int d^4x\int_0^1 ds{\it A}\left( x;s\Theta \right)
\Theta \left( x\right) =\int d^4xL_{eff}-W_{WZW}\eqno(7)
$$

where the Wess-Zumino-Witten term $W_{WZW}$ describes topological
properties of $g_5$ (and $U$ ) and is represented by a
five-dimensional integral with $x_5=s$. It is the analogue of
$W_{eff}$ for color that we are interested in.

\vskip 9pt

 {\large \bf 3. Color bosonization}

\vskip 9pt

 The basic color fields are the Yang-Mills field $V_\mu
\left( x\right) $ and the quark field $\bar \psi \left( x\right)
,\psi \left( x\right) $ with the Lagrangian

$$
L_\psi =\bar \psi \not D\left( V\right) \psi , \eqno(8)
$$

where $\not D\left( V\right) $ is the Dirac operator for massless
quarks with the Yang-Mills field $V_\mu $. There are no dynamical
axial vector field: $A_\mu =0$.

We consider the vacuum functional $Z$ for the system quarks +
gluons as being always in the presence of the color chiral field
$U\left( x\right) $ describing local chiral degrees of freedom of
quarks $\bar \psi ,\psi $ and resulting in replacement of $Z_\psi
\left[ V\right] $ by $Z_\psi \left[ V^U,A^U\right] $, where $V^U$
and $A^U$ are vector and axial vector fields arising in the Dirac
operator $\not D\left( V\right) \rightarrow \not D\left(
V^U,A^U\right) $ from the gluonic field $V_\mu $ due to chiral
rotation

$$
Z=\int d\mu _V\{\exp i\int dxL_{YM}\left( V\right) \}Z_\psi \left[
V^U,A^U\right] \eqno(9)
$$

The vacuum functional $Z$ depends on color degrees of freedom of
quarks and gluons. The gluon measure $d\mu _V$ includes only
vector color degrees, while the quark functional $Z_\psi $
contains both vector and chiral color degrees in the quark measure
$d\bar \psi d\psi $. Explicitly Z$_\psi $ depends only on gluonic
field $V_\lambda $. Under transformations of the color gauge group
$SU \left( 2\right)_c=SU\left( 2\right) _{L +R}$ the vacuum
functional is invariant,$\delta $Z=0. The existence of chiral
anomaly means that $\delta Z\neq 0$ under chiral transformations
belonging to the coset $G_{LR}/SU\left( 2\right) _{L+R}$, because
of non-invariance of the measure\cite{fujikawa}.

While in the flavor case $V_\mu $ and $U$ are always independent
variables, in the color case two types of questions are posiible:

(a) what is an action for the chiral field in a given gluon field.
For example, what is an action for chiral soliton in color vacuum
field \cite{A+N}. Total number of variables should not exceed that
of dynamical gluon field. This question is of the same type as in
the flavor case, and an action is given by an expression (7) for
flavor one.

(b) What is an anomalous action for color variables taking into
account that a pair $\left( V^U,A^U\right) $ should contain the
same number of variables as $V$. Then an initial path integral is
$Z_U$ in (9) with all color variables $\left( V^U,A^U\right) $ of
left-right group shown explicitly. In this case, an anomalous
action is defined by the expression, which formally is of opposite
sign compared with the flavor case (7). The ''bosonized'', or
anomalous action is defined by the expression

$$
W_{bose}\left[ V^U,A^U\right] =-i(\ln Z_\psi \left[ V^U,A^U\right]
-\ln Z_\psi \left[ V,0\right] ) , \eqno(10)
$$

which includes in general two different actions: a topological
$W_{WZW}$ and a non-topological $W_{an}$ ones. For SU(2)
$W_{WZW}=0$. For SU(3) it is the most interesting part.

We consider one loop approximation for gluons in the background
gauge in absence of external vector fields. Then $V_\mu $ will be
a classical (background) field for gluons. There is no background
axial vector field $A_\mu $. After quark color chiral
transformation with the chiral field $U$ , we get from $V_\mu $ a
vector matrix $W_\mu ^U$ containing both $V_\mu ^U$ and $A_\mu ^U$

$$
W_\mu ^U=V_{\mu a}^Ut_a+A_{\mu a}^Ut_{5a}\\V_\mu ^U=\frac 12\left(
U^{+}V_\mu U+V_\mu +U^{+}\partial _\mu U\right) ,$$
$$A_\mu
^U=\frac 12U^{+}D_\mu U ,\eqno(11)
$$

where $D_\mu $ is defined with the background field $V_\mu $. When
$U=1$ we return to the case, when $A^U=0$ and $W_\mu ^U=V_\mu $.
Note that a chirally rotated gauge field is just an extension of
an initial gauge field by an induced axial vector field: $V_\mu
^U=V_\mu +A_\mu ^U$.

Thus, in the color bosonization approach, there are three vectors
$V_\mu ,V_\nu ^U,A_\lambda ^U$ living in the common color space of
gluons and chiral color of quarks. They include two gauge fields
$V_\mu $ and $V_\mu ^U$, belonging to different vector-type
subgroups of left-right chiral group. $V_\mu $ transforms with
$L_a+R_a$ color generators, while $V_\mu ^U$ transforms with
generators $L_a^U+R_a$, where the left generator is additionally
rotated. We remind that the chiral field $U$ belongs to the
anomalous channel $\Theta $.

Because of chiral transformations, asymptopic constraints imposed on gluonic
field $V_\mu $ lead to constraints for $V_\mu ^U,A_\mu ^U$. It is usually
required for the dynamical $V_\mu $ that

$$
\int d^3xtrV_\mu V_\mu \prec \infty \eqno(12)
$$

We assume that this property is preserved by chiral
transformations. In view of orthogonality of $t_{5a}$ and $t_b$,
we have

$$
\int d^3x\{trA_\mu ^UA_\mu ^U\} \prec \infty \eqno(13)
$$

It means that asymptotically

$$
U^{+}D_\mu U\rightarrow 0,r\rightarrow \infty
$$

Consider the chiral field $U=\xi _L^2$, where $\xi _L$ is an
$SU\left( 2\right) _L$ rotation in the fundamental representation
with generators $\tau _a/2$

$$
\xi _L\left( x\right) =\exp \left( i\hat nF/2\right) ,0\leq F\leq 2\pi
$$
$$ U = \exp 2(i\hat nF/2),\hat n=n_a\tau _a,n_an_a=1 \eqno(14)$$
$$
U^{+}D_\mu U = i\hat n\partial _\mu F+i\frac 12D_\mu \hat n\sin
2F+\left[ \hat n,D_\mu \hat n\right] \frac 12\sin ^2F
$$

Then asymptotically at $r\rightarrow \infty $

$$
\partial _\mu F\rightarrow 0,D_\mu \hat n\rightarrow 0
$$

Within the chiral left-right group gauge fields $V_\mu $ and
$V_\mu ^U$ are associated with different SU(2)- subgroups. One may
conjecture that there is a finite region in the total color space,
where both fields are equivalent. Such a region should correspond
to restricted number of color degrees of freedom. A boundary of
this region, where number of variables is changed, will be
reflected in behaviour of the determinant $\det \left[ 1+R\left(
U\right) \right] $.

Consider a change of fields $A_\mu \left( x\right) \rightarrow
A_\mu ^{\prime }\left( x\right) =\frac 12\left( U^{+}\left(
x\right) A_\mu \left( x\right) U\left( x\right) +A_\mu \left(
x\right) \right) $, or

$$
A_{\mu a}^{\prime }=\frac 12\left( \delta _{ab}+R_{ab}\left(
U\right) \right) A_{\mu b},R_{ab}\left( U\right) =\frac 12tr\left(
\lambda _aU^{+}\lambda _bU\right) \eqno(15)
$$

where $R_{ab}\left( U\right) $ is the transformation $U$ in
adjoint representation. We get 3$\times 3$ matrix $R\left(
U\right) =R\left( \xi
_L^2\right) $ by replacing $SU\left( 2\right) $ generators $\tau _a/2$ with $%
SO(3)$ hermitian generators $O_a$

$$
R\left( U\right) =\exp iO_an_a2F=1+i\hat N\sin 2F+\hat N^2(\cos
2F-1) \eqno(16)
$$

The 3$\times 3$ matrix $\hat N=O_an_a$ has a property $\hat
N^3=\hat N$ , so that eigenvalues of $\hat N$ are equal to
+1,-1,0. It follows that $\det \frac 12\left( 1+R\left( U\right)
\right) =\frac 12\left( 1+\cos 2F\right) $. Thus, we have
singularities at points $x_s$, where $F\left( x_s\right) =\pi /2$
, $U_s\equiv U\left( x_s\right) =i\hat n\left( x_s\right) $ and $
R\left( U_s\right) =1-2\{\hat N\left( x_s\right) \}^2$. At
singularity two eigenvalues of the matrix $R\left( U\right) $
coincide. In the chiral color space, these singular points $x_s$
constitute a spherical surface of radius $ F\left( x_s\right) $ in
the anomalous channel (i.e. which is gauge equivalent to a region
of $\gamma _5\theta $ -parameter of the left-right group), where a
general gluonic field $V_\mu $ cannot be expressed in terms of
chirally dependent field $V_\mu ^U$. The transformation $V_\mu
\left[ x\right] \rightarrow \frac 12\left[ 1+R\left( U\right)
\right] V_\mu \left( x\right) $ may be induced by a global chiral
rotation, $\partial _\mu U=0$; thus, already a global chiral
rotation leads to a singular determinant.

Explicitly, using an expression $V_\mu ^U=V_\mu +A_\mu ^U$ we get in terms
of color vectors $\vec n,\vec V_\mu $

$$
\vec V_\mu ^U=\vec V_\mu \cos 2F-\frac 12\left[ \vec V_\mu ,\vec
n\right] \sin 2F-\vec n(\vec V_\mu ,\vec n)\sin ^2F+  $$
$$ i\vec
n\partial _\mu F- \frac 12\partial _\mu \vec n\sin 2F-\left[ \vec
n,\partial _\mu \vec n\right] \sin ^2F \eqno(17)
$$

so that at $F=\pi /2$ the field $V_\mu ^U$ looses structures represented by $%
\partial _\mu \vec n$ and $\left[ \vec V_\mu , \vec n\right] :$

$$
V_\mu ^U=\frac 12\left( \hat n\partial _\mu \hat n+\hat n\hat
V_\mu \hat n+V_\mu \right) ,F=\frac \pi 2  \eqno(18)
$$

while the matrix $\Delta =\frac 12\left( 1+R\left( U\right)
\right) $ reduces to $\Delta ^0=1-\hat N^2$ with matrix elements
$\Delta _{ab}^0=n_an_b,$ and determinant $\det \Delta ^0=0$. The
matrix $1-N^2$ is a projector on eigenvalue $\hat N^{\prime }=0$.
Thus, at $F=\pi /2$, only fields with $N^{\prime }=0$ are
essential for construction of singular free connections in color
space including chiral degrees of freedom introduced by $U$. In
fact, this conclusion follows directly from properties of $\Delta
$

$$
\left( \Delta \right) _{ab}n_b=n_a,(\Delta ^{-1})_{ab}n_b=(1-i\hat
N\tan F)_{ab}n_b=n_a  \eqno(19)
$$

It reflects (by construction of the chiral field $U=\exp i\hat nF$
) the fact that $U$ commutes locally with a function of $\hat n$.
In the case, when a relation $UV_\mu U^{+}=V_\mu $ helds locally,
we have $\det \Delta =1$.

Let us demonstrate, that it is possible to find such a gluon field
$V_\mu $ and such a chiral field $U$, that we have an invariance
relation $V_\mu =V_\mu ^U$. We represent $V_\mu $ in the form
$V_\mu =C_\mu \hat n+\frac 12U\partial _\mu U^{+}$, where $C_\mu $
is an abelian gauge field and check the relation $V_\mu ^U=\frac
12\left( UV_\mu U^{+}+V_\mu +U\partial _\mu U^{+}\right) =V_\mu $.
It can be satisfied if $U=i\hat n$ or $F=\pi /2$. We denote this
special field on the sphere $\Omega \left( \pi /2\right)$ by
$V_\mu ^\Omega $

$$
V_\mu ^\Omega =C_\mu \hat n+\frac 12\hat n\partial _\mu \hat n, $$

$$ A_\mu ^\Omega =\frac 12\left( UV_\mu ^\Omega U^{+}-V_\mu
+U\partial _\mu U^{+}\right) =0 \eqno(20)
$$

The chiral field $U$ and the gauge field $V_\mu $ contain now the
same unit color vector $\hat n$. Second term in $V_\mu ^\Omega $
satisfies separately the equivalence relation $V_\mu ^U=V_\mu $ .
The field $V_\mu ^\Omega $ is a basic field in the color
bosonization approach, because then an axial vector $A_\mu
^U\left( \Omega \right) =0$; both the Yang-Mills action
$I_{YM}\left( V^\Omega \right) $ and the quark integral $Z_\psi
\left[ V^U,A^U\right]$, which is in $\Omega $ just $Z_\psi \left[
V^\Omega ,0\right] $, depend on the same $V^\Omega $ only. There
is no color chiral anomaly -neither a topological one (for SU(3)),
nor of a non-topological type. The field $V_\mu ^\Omega $ can be
obtained by chiral transformation from the simplest vector field,
namely, from an abelian field $V_\mu ^0=C_\mu \hat n$. Also,
$V_\mu ^\Omega $ is invariant under the gauge transformation with
the chiral field $U=i\hat n$ and under chiral transformation

$$
V_\mu ^\Omega =(V_\mu ^0)^U=\frac 12(\hat nV_\mu ^0\hat n+V_\mu
^0+\hat n\partial \hat n),V_{\mu ^{}}^\Omega =\hat nV_\mu ^\Omega
\hat n+\hat n\partial _\mu \hat n
$$

The corresponding Yang-Mills field strength $ V_{\mu \nu }^\Omega
$ is

$$
V_{\mu \nu }^\Omega =C_{\mu \nu }\hat n+\frac 14\left[ \partial
_\mu \hat n,\partial _\nu \hat n\right] \eqno(21)
$$

The field $V_\mu ^\Omega $ depends on four degrees of freedom,
instead of required 6 degrees in the case of SU(2). The field
$V_\mu ^\Omega$ was introduced as a starting point of $n-$model
\cite{Faddeev75}(Faddeev) and "Restricted gauge theory"
\cite{Cho}(Cho).

Fixing $det \Delta =1$ corresponds to excluding the Cartan mode
$\exp i\tau _3F$ from the chiral field $U=\exp i\hat nF$ . In
terms of $SU\left( 2\right) _L\times SU\left( 2\right) _R$
generators $\tau _a/2,\gamma _5\tau _b/2$ , it means that we
rotate $\hat n$ to $\tau _3$ and then fix $\gamma _5\tau _3-$
parameter $F$.

\vskip 9pt

{\large \bf 4. Gluonic chiral anomaly for basic field $V_\mu
^\Omega $, as a background field, and kinetic term for $n$ -
field}

\vskip 9pt

A quantum expression for the Jacobian of $V_\mu \rightarrow V_\mu ^U$
transformation is given by the chiral anomaly.

Consider the path integral $Z\left[ U\right] $ in the background
gauge, where the chiral field $U$ is in region $\Omega $ of the
chiral color space, i.e in the space of chiral parameters $\theta
\hat =\hat n$. We have a gluonic background field $V_\mu ^\Omega $
in gluonic part of $Z\left[ U\right] $ and a background field
$V_\mu ^U=V_\mu +A_\mu ^U$ in fermionic path integral $Z_\psi
\left[ V_\mu^U,A_\mu ^U\right] $

$$
Z\left[ U\right] =\int d\mu _Q\exp iL_{eff}\left( V+Q\right)
Z_\psi \left[ V^U,A^U\right] \eqno(22)
$$

where $Q_\mu $ is a quantum field and $G_\mu =V_\mu +Q_\mu $ is a
dynamical gluon field. In $\Omega $, these background fields are
equal, while $A_\mu ^U=0$, so that QCD in $\Omega $ is associated
with the path integral

$$
Z^\Omega =\int_\Omega d\mu _Q\exp iL_{eff}\left( V\left( \Omega
\right) +Q\right) Z_\psi \left[ V\left( \Omega \right) ,0\right]
\eqno(23)
$$

extended over color vector variables $Q_\mu $. In the case of
bosonization, the anomaluos action resulting from integration of
non-abelian anomaly of fermionic path integral $Z_\psi $ is a
function of axial vector fields $A_\mu ^U$. Thus, in $\Omega $
there is no anomalous color action arising from $Z_\psi $.

In the complete color space, the anomalous action is composed of
two pieces coming from the gluonic part and from bosonization. Let
us retain only the gluonic part. Then we should start by
considering the path integral $Z_{an}\left[ U\right] $ over
non-chirally transformed gluons $G_\mu $ and ghosts (measure
$\left[ dG\right] $ ) with an effective lagrangian $L_{eff}\left(
G^U\right) $ for a chirally transformed field $G^U$ and ghosts

$$
Z_{an}\left[ U\right] =\int \left[ dG\right] \{\exp i\int
dxL_{eff}\left( G^U\right) \} \eqno(24)
$$

The effective lagrangian $L_{eff}$ includes the Yang-Mills part
$L_{YM}$, gauge fixing term $L_{GF}$ and ghost lagrangian
$L_{FP}$. The ratio $Z_{an}\left[ U\right] /Z_{an}\left[
U=1\right] $ gives us the Jacobian of $(V\rightarrow V^U)$ -
transformation.

We are interested in one-loop part of $L_{eff}\left( G^U\right) $ in the
general case for the field $G_\mu $ and an arbitrary $U$. We introduce a
background field $B_\mu \left( x\right) $ and write

$$
G_\mu =B_\mu +Q_\mu \eqno(25)
$$

where $B_\mu $ transforms inhomogeneously

$$B_\mu ^{\prime
}=gB_\mu g^{+}+g\partial _\mu g^{+}$$

under gauge transformations $g\left( x\right) $ , while $Q_\mu
^{\prime }=gQ_\mu g^{+}$.

Due to linear relation between $Q_\mu ^U$ and $Q_\mu $, one-loop
part $L_{1L}\left( Q\right) $ of the effective lagrangian
$L_{eff}\left( G^U\right) $ in $Q_\mu $ follows immediately from
one-loop part $L_{1L}\left( Q^U\right) =Q_\mu ^UK_{\mu \nu }\left(
B^U\right) Q_\nu ^U $ in $ Q_\mu ^U$

$$
L_{1L}\left( Q\right) =Q\Delta ^TK\left( B^U\right) \Delta Q
\eqno(26)
$$

where matrix $K_{\mu \nu }^{ab}\left( B^U\right) $ depends on
rotated background field $B_\mu ^U$, and the matrix $\Delta
_{ab}=\frac 12\left( 1+R\left( U\right) \right) _{ab}$ includes a
color rotation $R_{ab}=\frac 12 tr\left( U\tau _aU^{+}\tau
_b\right) $, $\Delta ^T$ is transposed of $ \Delta $. For $K_{\mu
\nu }\left( B^U\right) =D^2\left( B^U\right) \delta _{\mu \nu
}-2B_{\mu \nu }^U$ we have

$$
D_\lambda \left( B^U\right) = D_\lambda \left( B+A^U\right)
=D_\lambda \left( B\right) +\left[ A_\lambda ^U,*\right]
$$
$$
B_{\mu \nu }^U =(B+A^U)_{\mu \nu }= $$ $$ B_{\mu \nu }+D_\mu
\left( B\right) A_\nu ^U-D_\nu \left( B\right) A_\mu ^U+\left[
A_\mu ^U,A_\nu ^U\right]\eqno(27)
$$

Thus, the one-loop quantum corrections to the Jacobian $J\left(
\partial G_\mu ^U/\partial G_\mu \right) $ arising from integration
over variables $Q_\mu $ with background field $B_\mu $ are given
by

$$
Tr\ln \{\Delta ^TK\left( B^U\right) \Delta \}_{Reg} \eqno(28)
$$

where $Reg$ denotes regularization.

The special case of a background field in $\Omega $ corresponds to
$B_\mu =V_\mu =V_\mu ^\Omega $, $A_\mu ^U=0$ and $\Delta =1-N^2$;
so that $\Delta $ becomes a projector on $N^{\prime }=0$ state.
Then the matrix $\Delta ^T\Delta =M=1-N^2$ is also the projector,
while the one-loop expression for gluonic chiral anomalous action
reduces to the standard one-loop gluodynamics taken with $\Delta
-$ projectors from both sides

$$
Tr\ln \{\Delta ^T[p^2+D^2+2pD-2V]\Delta \}_{Reg}+\left(
ghosts\right) \eqno(29)
$$

However, the term $\ln \Delta ^T\Delta $ is not defined by such a
direct replacement $\Delta \rightarrow 1-N^2$. In order to give
meaning to expressions like $\ln \Delta $ and be sure that we can
drop $\Delta ^{\prime }$s from $\Delta p^2\Delta $ term in (29),
we need at the first step to retain in $\Delta $ the terms with
projector N$^2$ on states with $N^{\prime }=\pm 1$. Then terms
quadratic in $\partial _\mu N^2$ will give us the kinetic
lagrangian
$$-\Lambda ^2\frac 12trD_\mu N^2N^2D_\mu N^2\rightarrow
\frac{\Lambda ^2}2\partial _\mu n_a\partial _\mu n_a $$, where
$\Lambda $ is a cut-off.

\vskip 9pt

{\large \bf 5. QCD-SU(2)$_c$ at low energies: gauge field and the
effective gluonic action}

\vskip 9pt

The field $V_\mu ^\Omega $ is a common part of initial gauge field
$V_\mu $ and chirally rotated version $V_\mu ^U$. In left-right
group without dynamical $A_\mu $ these gauge fields are
interrelated by $V_\mu =V_\mu ^\Omega -A_\mu ^U$. Chiral rotation
of quarks $i\hat n$ transforms $Z\psi [V^\Omega -A^U,0]$ into
$Z_\mu [V^\Omega ,A^U]$, acting as a shift operator. $A_\mu ^U$
should anticommute with $\hat n$. Denoting $A_\mu ^U=-X_\mu $ we
come to the decomposition for the QCD gauge field

$$
V_\mu =V_\mu ^\Omega +X_\mu \eqno(30)
$$

This decomposition for the QCD gauge field was discussed by Cho
\cite{Cho} in different approach.

For color bosonization approach anticommutativity relation $\{\hat
n,X_\mu \}=0$ is essential. Due to this property of $X_\mu $, a
chirally rotated gauge field $V_\mu {}^U=(V_\mu ^\Omega +X_\mu
)^U=V_\mu ^\Omega $ is independent of $X_\mu $, while the axial
field $A_\mu ^U$ picks up the value $A_\mu ^U=-X_\mu $. Thus, this
expression for a gauge field $V_\mu $ is, at the same time, a
decomposition of a gauge field into chirally rotated vector part
$V_\mu ^U=V_\mu ^\Omega $ and chirally rotated axial vector part
$A_\mu ^U=-X_\mu $. In this form, it is explicitly seen that the
gluon sector and the quark sector are built on the same color
variables, and theYang-Mills action $I_{YM}\left( V\right) $ and
bosonization part $Z_\psi \left[ V^\Omega ,-X\right] $ contain the
same set of background fields $ (V_\mu ^\Omega ,X_\mu )$. In the
SU(2) case, an axial field $A_\mu ^U=-X_\mu $ leads to a
non-topological chiral anomaly of the quark integral $Z_\psi $. In
the case SU(3) the chiral anomaly will include also a topological
term. Thus, in this decomposition of $V_\mu $, the field $X_\mu $
is responsible for the chiral anomaly, and consequently, for a
bosonization action. Such an action together with the Yang-Mills
action and kinetic term will determine low energy color dynamics.

The color bosonization action $W_{an}$ can be written in analogy
with the flavor case \cite{AANN}. In our notations, the
non-topological part of $W_{an}$ corresponds to the following
Lagrangian $L_{an}=L_{+}\left( V_\mu ^\Omega ,-X_\mu \right)
-L_{+}\left( V_\mu ^\Omega +X_\mu ,0\right) $ in the Minkowski
space

$$
L_{an}=\frac{\Lambda ^2}{4\pi ^2}trX_\mu ^2-\frac 1{12\pi
^2}tr\{\frac 14\left( V_{\mu \nu }^\Omega \right) ^2+X_\mu V_{\mu
\nu }^\Omega X_\nu - $$ $$ \frac 12\left[ D_\mu ^\Omega ,X_\mu
\right] ^2- \frac 14\left[ X_\mu ,X_\nu \right] ^2+\left( X_\mu
^2\right) ^2\}+ \frac 1{48\pi ^2}tr \left( V^\Omega +X\right)
_{\mu \nu }^2  \eqno(31)
$$

where $D^\Omega $ contains the field $V_\mu ^\Omega $, while

$$V_{\mu \nu }^\Omega =C_{\mu \nu }\hat n+\frac 14\left[ \partial
_\mu \hat n,\partial _\nu \hat n\right] ^{}$$

is the field strength of $V_\mu ^\Omega $.

The Yang-Mills Lagrangian for $V_\mu =V_\mu ^\Omega +X_\mu $ is
given by

$$
L_{YM}=L_{YM}^\Omega +\frac 1{2g^2}tr\{ \left( D_\mu ^\Omega X_\nu
-D_\nu ^\Omega X_\mu \right) ^2+ \left[ X_\mu ,X_\nu \right]
^2+2V_{\mu \nu }^\Omega \left[ X_\mu ,X_\nu \right] \} \eqno(32)
$$

The effective $SU(2)$ gluonic Lagrangian in variables ($V_\mu
^\Omega ,X_\nu )$ is

$$
{\it L}=L_{an}+L_{YM}+\frac{\Lambda ^2}{4\pi ^2}tr(\partial _\mu
\hat n)^2=L_{YM}^\Omega +\frac{\Lambda ^2}{4\pi ^2}tr(\partial
_\mu \hat n)^2+ $$
$$T+P+ (\frac 1{24\pi ^2}+\frac 1{g^2})tr\{V_{\mu
\nu }^\Omega \left[ X_\mu ,X_\nu \right] \} \eqno(33)
$$

where $T$ is the kinetic term for $X_\mu $ and ($-P)$ is the potential

$$
T=(\frac 1{48\pi ^2}+\frac 1{2g^2})tr\{\left( D_\mu ^\Omega X_\nu
-D_\nu ^\Omega X_\mu \right) ^2\}+  \frac 1{24\pi ^2}tr\left[
D_\mu ^\Omega ,X_\mu \right] ^2 \eqno(34)
$$ $$
P=\frac{\Lambda ^2}{4\pi ^2}trX_\mu ^2-\frac 1{12\pi ^2}tr(X_\mu
^2)^2+ (\frac 1{24\pi ^2}+\frac 1{2g^2})tr\left[ X_\mu ,X_\nu
\right] ^2 \eqno(35)
$$

It follows  that $trX_\mu ^2$ can form a gauge invariant
condensate $\left( trX_\mu ^2\right) _0=-g^2\sigma /2$ as a
minimum of $-P$ , and a mass appears. Denote a hermitian vacuum
field by $\phi _\mu ^a$, so that  $\sigma =(\phi _\mu ^a\phi _\mu
^a)$ and $X_\mu =(\phi _\mu ^a+Y_{\mu ^{}}^a)\tau _a/2i$. Then
$\sigma =-9\Lambda ^2/(7g^2+48\pi ^2)$ and $m_Y^2=-\frac 13\sigma
b_3$ , where $b_3=(g^2+12\pi ^2)/24\pi ^2g^2$. The condensate
$\sigma $ is negative, and the vacuum field $\phi _\mu ^a$ is
space-like.

The last term in $T$ should be analysed together with the gauge
condition for $V_\mu $. In the flavor case, the term $\left(
-P\right) $ without $tr\{(X_\mu ^2)^2$ corresponds to the Skyrme
Lagrangian. The lagrangian for the n-field \cite{Faddeev75} is
contained in the first two terms of ${\it L}$.

The Lagrangian of \cite{Cho} is $L_{YM}(V^\Omega +X)$. New terms
are contained in $L_{an}$; they are partly built on structures
already existing in $L_{YM}$, but with different coefficients.
Note, that an expression $tr\left[ D_\mu ^\Omega ,X_\mu \right]
^2$ in color $L_{an}$ looks as a standard gauge condition term,
while in the flavor case $ \left [ D_\mu A_\mu \right ]^2$ leads
to ghosts. Last term in $T$ and first two terms in $P$ are quite
new; they are specific for bosonization. To get more insight into
meaning of $L_{an}$ we need to assume a definite representation
for $X_\mu$. Investigation of the effective Lagrangian is the next
step of bosonization approach.

\vskip 12pt

{\bf Acknowledgments}

We are grateful to Dmitry Vassilevich for interesting discussions.

\end{document}